\documentclass[aps,prl,twocolumn,showpacs,superscriptaddress,groupedaddress]{revtex4}  
\usepackage{graphicx}  
\usepackage{dcolumn}   
\usepackage{bm}        
\usepackage{amssymb}   
\usepackage{amsmath}

\usepackage{color}

\begin{document}



\title{First Online Mass Measurements of Isobar Chains via Multi-Reflection Time-of-Flight Mass Spectrograph Coupled with GARIS-II}
\date{\today}

\author{P.~Schury}
\affiliation{Institute of Physics, University of Tsukuba, Ibaraki 305-8571, Japan}
\affiliation{RIKEN Nishina Center for Accelerator-Based Science, Wako, Saitama 351-0198, Japan}
\author{M.~Wada}
\affiliation{Institute of Particle and Nuclear Studies (IPNS), High Energy Accelerator Research Organization (KEK), Ibaraki 305-0801, Japan}
\affiliation{RIKEN Nishina Center for Accelerator-Based Science, Wako, Saitama 351-0198, Japan}
\author{Y.~Ito}
\affiliation{RIKEN Nishina Center for Accelerator-Based Science, Wako, Saitama 351-0198, Japan}
\author{D.~Kaji}
\affiliation{RIKEN Nishina Center for Accelerator-Based Science, Wako, Saitama 351-0198, Japan}
\author{P.-A. S\"{o}derstr\"{o}m}
\affiliation{RIKEN Nishina Center for Accelerator-Based Science, Wako, Saitama 351-0198, Japan}
\author{A.~Takamine}
\affiliation{RIKEN Nishina Center for Accelerator-Based Science, Wako, Saitama 351-0198, Japan}
\author{F.~Arai}
\affiliation{RIKEN Nishina Center for Accelerator-Based Science, Wako, Saitama 351-0198, Japan}
\author{H.~Haba}
\affiliation{RIKEN Nishina Center for Accelerator-Based Science, Wako, Saitama 351-0198, Japan}
\author{S.~Jeong}
\affiliation{Institute of Particle and Nuclear Studies (IPNS), High Energy Accelerator Research Organization (KEK), Ibaraki 305-0801, Japan}
\author{S.~Kimura}
\affiliation{RIKEN Nishina Center for Accelerator-Based Science, Wako, Saitama 351-0198, Japan}
\author{H.~Koura}
\affiliation{Advanced Science Research Center, Japan Atomic Energy Agency, Ibaraki 319-1195, Japan}
\author{H.~Miyatake}
\affiliation{Institute of Particle and Nuclear Studies (IPNS), High Energy Accelerator Research Organization (KEK), Ibaraki 305-0801, Japan}
\author{K.~Morimoto}
\affiliation{RIKEN Nishina Center for Accelerator-Based Science, Wako, Saitama 351-0198, Japan}
\author{K.~Morita}
\affiliation{RIKEN Nishina Center for Accelerator-Based Science, Wako, Saitama 351-0198, Japan}
\affiliation{Kyushu University, Hakozaki, Higashi-ku, Fukuoka 812-8581, Japan}
\author{A.~Ozawa}
\affiliation{Institute of Physics, University of Tsukuba, Ibaraki 305-8571, Japan}
\author{M.~Reponen}
\affiliation{RIKEN Nishina Center for Accelerator-Based Science, Wako, Saitama 351-0198, Japan}
\author{T.~Sonoda}
\affiliation{RIKEN Nishina Center for Accelerator-Based Science, Wako, Saitama 351-0198, Japan}
\author{T.~Tanaka}
\affiliation{RIKEN Nishina Center for Accelerator-Based Science, Wako, Saitama 351-0198, Japan}
\affiliation{Kyushu University, Hakozaki, Higashi-ku, Fukuoka 812-8581, Japan}
\author{H.~Wollnik}
\affiliation{New Mexico State University, Las Cruces, NM 88001, USA}

\begin{abstract}
Using a mulit-reflection time-of-flight mass spectrograph (MRTOF-MS) located after a gas cell coupled with the gas-filled recoil ion separator GARIS-II, the masses of several heavy nuclei have been directly and precisely measured.  The nuclei were produced via fusion-evaporation reactions and separated from projectile-like and target-like particles using GARIS-II before being stopped in a helium-filled gas cell.  Time-of-flight spectra for three isobar chains, $^{205}$Fr\textemdash $^{205}$Rn\textemdash $^{205}$At\textemdash  $^{205}$Po, $^{206}$Fr\textemdash $^{206}$Rn\textemdash $^{206}$At and $^{201}$Rn\textemdash $^{201}$At\textemdash $^{201}$Po\textemdash $^{201}$Bi, were observed.  Precision atomic mass values were determined for $^{205, 206}$Fr, $^{201}$At, and $^{201}$Po.  

\end{abstract}

\pacs{}
\maketitle


In recent years the multi-reflection time-of-flight mass spectrograph (MRTOF-MS), first proposed by Wollnik~\cite{Wollnik90} more than 20 years ago, have begun to make an impact in nuclear physics as isobar separators \cite{Wolf2013, Dickel2015} and for use in precision mass measurements \cite{Wienholtz2013, Ito2013, Ishida2005}.  We hope to be able to use the MRTOF-MS to eventually change from a paradigm of identifying transactinide isotopes by $\alpha$-decay chain to one of identification by mass determination.  Doing so will reduce uncertainty in identification, especially in the region of hot-fusion where $\alpha$-decay chains terminate in spontaneous fission before reaching well-known nuclei, which has bottlenecked acceptance of elements 113, 115, 117 and beyond \cite{IUPAC}.

The MRTOF-MS is the ideal tool for such a paradigm shift, being well-suited for low-yield, heavy, and short-lived nuclei.  It achieves mass resolving powers $R_{\textrm{m}} >$ 100,000 with flight times shorter than 10~ms for even the heaviest nuclei \cite{Schury2014}.  Additionally, it is a true spectrograph \textemdash~as opposed to a spectrometer \textemdash~making it capable of mass determinations with, in principle, as few as one detected ion.
Owing to its spectrographic nature, the MRTOF-MS is able to simultaneously measure several species with high-precision.  This capability has until now been limited to storage rings.  As we have demonstrated \cite{Ito2013, Schury2014}, and will herein further demonstrate, the MRTOF-MS allows for a considerably less complex data analysis than that used for storage rings.

As the first step toward mass spectrographic identification of SHE, we have installed a gas cell connected to an MRTOF-MS after the gas-filled recoil ion separator GARIS-II \cite{GARIS-II}.  The system is described in some detail elsewhere \cite{SchuryEMIS2015}.  We have used this system to initially perform mass measurements with fusion-evaporation reaction products lighter than Uranium, the masses of some of which have not previously been directly measured.  In these measurements we demonstrate the ability of the MRTOF-MS to precisely determine the masses of several isotopes simultaneously and to do so in some cases with less than 10 detected ions.


A 1.5~p$\mu$A beam of $^{40}$Ar$^{11+}$ at 4.825 MeV/$u$ was provided by the RIKEN heavy-ion linear accelerator RILAC.  The beam impinged upon a rotating target wheel with 16 target windows.  The target wheel comprised 4 windows of $^{165}_{~67}$Ho  with a thickness of $\sim$0.14 mg/cm$^2$ and 12 windows of $^{169}_{~69}$Tm with a thickness of $\sim$0.29 mg/cm$^2$.  The $^{165}$Ho and $^{169}$Tm targets were prepared using sputtering and electro-deposition methods, respectively, on 3~$\mu$m Ti backing foil. A rotating shadow wheel ensured the beam could only impinge on one type of target at a time \cite{Kaji2015}.  

When the projectile beam interacts with the target, the projectile and target nuclei can fuse to form a compound nucleus in an excited state.  The compound nucleus will then rapidly de-excite by particle emission, typically multiple neutrons ($x$n), a proton and multiple neutrons (p$x$n), or an $\alpha$ particle and multiple neutrons ($\alpha$$x$n) are emitted.  The remaining evaporation product, in this case, will pass out from the target with an energy of $\sim$30~MeV and an energy spread of several percent.  

The products were separated in-flight from projectiles and target-like particles using GARIS-II.  The separator was filled with Helium gas at 73~Pa.  Magnetic rigidity for each reaction product was set to be 1.669 Tm.  The ions passed through an exit window of 0.5~$\mu$m thick Mylar upon leaving GARIS-II.  

In order for these radioactive ions (RI) to be captured in an ion trap, they were first stopped and thermalized in a Helium-filled gas cell installed in the focal plane chamber following GARIS-II.  A degrader was placed between the GARIS-II exit window and the gas cell's 2.5~$\mu$m thick Mylar entrance window.  This degrader consisted of a 4~$\mu$m thick Mylar foil which could be rotated up to 45$^\circ$ to adjust the effective thickness. The gas cell was pressurized to $\approx$10~kPa. 

A flat array of silicon PIN diodes could be inserted between the GARIS-II exit window and the rotatable degrader. From the $\alpha$-decay spectrum we could identity and determine the rates of incoming ions. 

Ions were quickly extracted from the gas cell by use of an axial DC gradient and a circular traveling wave RF carpet \cite{Arai2014}.  The carpet is 8~cm in diameter with 80~$\mu$m wires with 80~$\mu$m spacing between them, and a 320~$\mu$m diameter exit orifice, which provides an excellent differential pumping barrier.  As ions left the gas cell via the small exit orifice, they were transported via RF-multipole ion guides to a triplet of RF quadrupole ion traps, similar to that reported in \cite{Ito2013Trap}, where they accumulated and cooled.  The ions were then orthogonally ejected from the central trap, accelerated by a pulsed drift tube to 1.5~keV and transported 2.5~m to a second pulsed drift tube and deceleration optics, followed by a second set of RF-quadrupole traps wherein the ions were recaptured.  Just prior to the second pulsed drift tube was a Bradbury-Neilson gate \cite{BNG} capable of suppressing ions differing in $A/q$ from the desired species with mass resolving power $\sim$100.  The central trap in the second trap triplet served as the final MRTOF preparation trap.  After cooling in the final trap, ions were again orthogonally ejected and entered the MRTOF-MS, wherein they underwent a predetermined number of reflections before being released to a multichannel plate (MCP) detector.  The ejection from the final trap served as a start signal for a time-to-digital converter (TDC) \cite{OurTDC}, while the signal on the MCP served as the TDC stop signal.  From accumulation in the first trap to  detection at the MCP, the cycle was 30~ms. 

The MRTOF-MS was tuned such that it achieved a maximum mass resolving power of $R_\textrm{m}$$\approx$150\,000 at $n$=148~laps.  At $n$=148 the mass bandwidth \cite{SchuryIJMS2014} is 1.34\%.  As such, in the mass region we investigated ions differing from the species of interest by more than $\pm$1.3~$u/q$ will appear in the ToF spectrum having made a different number of laps than the ions of interest and could by chance coincide in ToF near to the ions of interest.  To avoid mistaken identifications, we made all measurements at $n$=147 and $n$=148 laps; any peaks which do not appear at both numbers of laps may represent contaminants making a different number of laps.


The times-of-flight were determined by least-squares fitting of the spectral peaks to an exponential-Gaussian hybrid function \cite{EGH, Schury2014} using the MPFIT package \cite{MPFIT}.  In cases with $N$ simultaneous isobaric spectral peaks, the fitting function was a sum of $N$ exponential-Gaussian hybrid functions each having the same width and exponential decay rate.  The width and exponential decay rate were determined by scaling from the optimal fit parameters found for the $^{133}$Cs$^+$ reference in each measurement.

Atomic masses were calculated from reference measurements of $^{133}$Cs$^+$ in a single-reference analysis as described in \cite{Ito2013}.   We made use of a concomitant referencing method, to be described in detail separately, wherein online measurements and reference measurements are made sequentially in each cycle.  The multi-trap system allows one species to accumulate while the other is being analyzed, making the duty cycle $\approx$100\% for concomitant referencing.  The absolute time-of-flight for each detected ion was recorded with a precision of 100~ps, along with the cycle number.

Each data run was of duration longer than 30~minutes.  During measurements of extended duration the spectral peaks may drift due to thermal expansion of the reflection chamber and high-voltage power supply instabilities on the level of parts per million \cite{Schury2014}.  In the concomitant referencing method, the reference and measurement to drift together, allowing correction for such drifting.  The data were divided, based on cycle number, into $i$ subsets such that in each subset the reference spectral peak could be fit with a relative precision of $\delta t/t<$2$\times$10$^{-7}$.  The time-of-flight for each ion was then adjusted according to Eq.~\ref{egAdjust}. 

\begin{equation}
\label{egAdjust}
t^\textrm{corrected} = t\cdot t^\textrm{ref}_i/t^\textrm{ref}_{i=0}
\end{equation}

The masses were determined using Eq.~\ref{eqRatio}:
\begin{equation}
\label{eqRatio}
m = \rho\cdot m_\textrm{ref} = \bigg(\frac{t-t_0}{t_\textrm{ref}-t_0}\bigg)^2\cdot m_\textrm{ref}
\end{equation} 
where $t_0$ is the delay between the TDC start signal (which also triggers ejection from the MRTOF-MS preparation trap) and the actual ejection from the MRTOF-MS preparation trap.  Using an oscilloscope, the delay between the trap ejection trigger and the the actual trap ejection was measured to be $\approx$40~ns with a rise time of $\approx$10~ns, leading us to adopt $t_0$=45(5)~ns in our analysis.



\begin{figure}[th]
	\centering
	 \includegraphics[width = 3in]{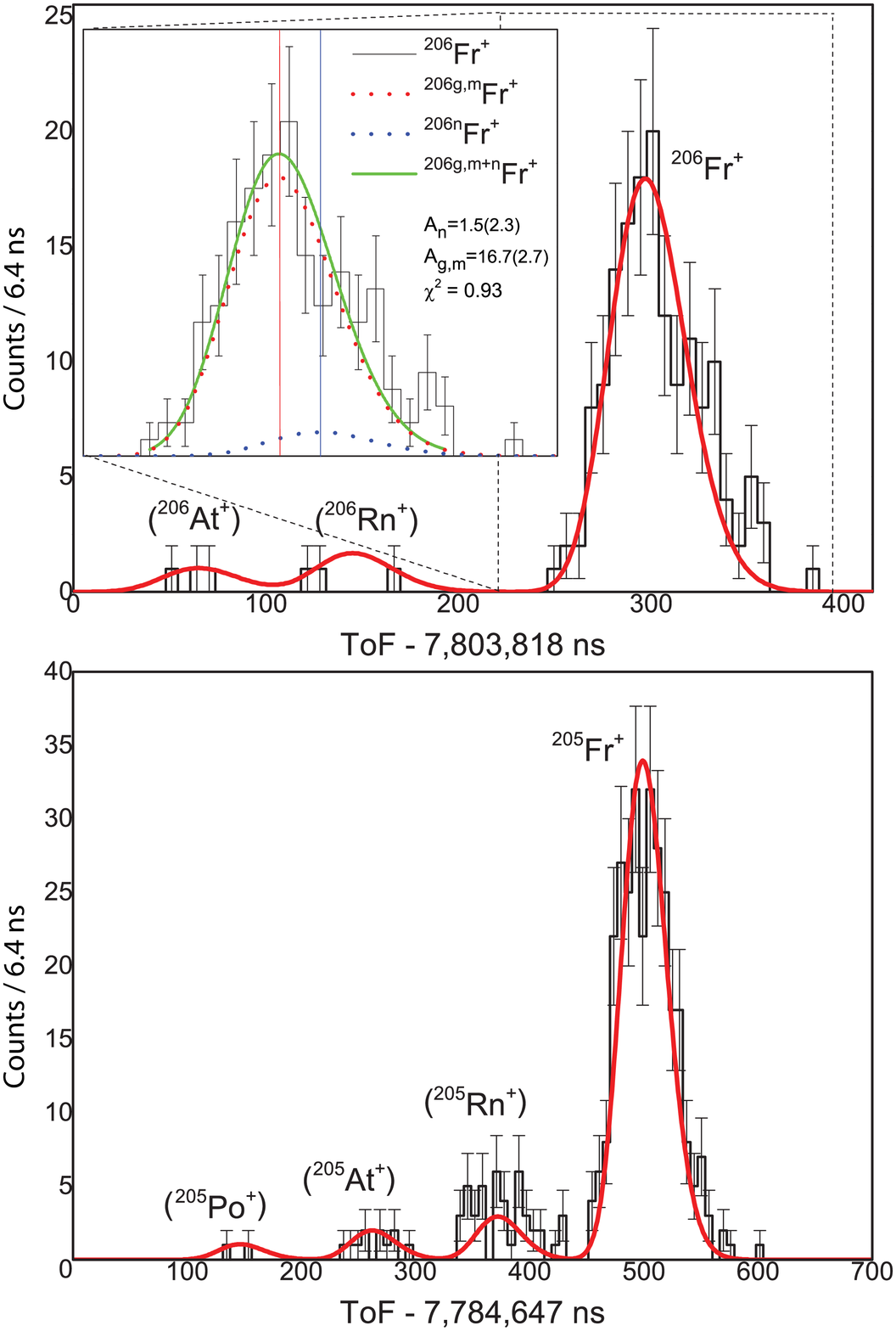} 
	\caption{\label{figFrSpectrum} Time-of-flight spectra observed using $^{169}$Tm target.  Ions made n=148 laps in the MRTOF-MS.  See text for details of parenthetical identifications.   \vspace{-5 mm}} 
\end{figure}

\begin{figure}[t]
	\centering
	\includegraphics[width = 3in]{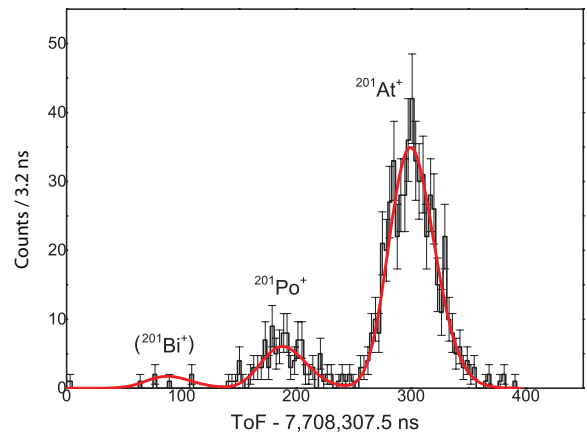}
	\caption{\label{figAtSpectrum} Time-of-flight spectrum observed using $^{165}$Ho target.  Ions made n=148 laps in the MRTOF-MS.  See text for details of parenthetical identification.  \vspace{-5 mm}}
\end{figure}

Using the $^{169}$Tm target, we simultaneously observed the 3n and 4n evaporation channel products $^{205,206}$Fr$^+$, as well as the p2n and p3n evaporation channel products $^{205, 206}$Rn$^+$.  In the spectrum at $n$=148 laps, ions were also detected which were consistent with the $\beta$-decay daughter and granddaughter of $^{205}$Rn, $^{205}$At$^+$ and $^{205}$Po$^+$ along with the $\beta$-decay daughter of $^{206}$Rn, $^{206}$At$^+$ (Fig.~\ref{figFrSpectrum}).  Using the $^{165}$Ho target, it was possible to simultaneously observe the 4n evaporation channel product $^{201}$At$^+$ and its $\beta$-decay daughter and granddaughter, $^{201}$Po$^+$ and $^{201}$Bi$^+$ (Fig.~\ref{figAtSpectrum}).  


The spectral peaks identified in Figs.~\ref{figFrSpectrum} and \ref{figAtSpectrum} as $^{205,206}$Rn, $^{206}$At, $^{205}$At, $^{205}$Po, and $^{201}$Bi$^+$ were not observed in the $n$=147 laps spectra.  While this is likely a result of shorter data runs for $n$=147~laps, it could also indicate that these spectral peaks correspond to non-isobaric contaminants making $n\ne$148~laps.  However, we can identify them with a high degree of confidence using the techniques described in \cite{SchuryWideband}. First, we calculated the masses that non-isobaric contaminants would need to have in order to be detected at the observed times-of-flight, limiting the possible mass range to $\pm$20~u due to the use of the Bradbury-Neilson gate.  In each case we found 4 candidate mass values within 0.15~u of an integer mass number.  We then calculated the times-of-flight where each of these candidates would be observed in the $n$=147 spectra.  For the $A/q$=205 and the $A/q$=201 species, no conjugate spectral peaks were observed in the $n$=147 spectra within 10~ppm of the calculated times-of-flight.  For the $A/q$=206 species, a spectral peak conjugate to $^{206}$Rn$^+$ and of similar intensity to that observed at $n$=148~laps, corresponding to $A/q$=187.921~u, was observed in the $n$=147 spectrum at $\approx$3~ppm from the calculated time-of-flight; no such conjugate spectral peak was observed for $^{206}$At$^+$.  Out of an excess of caution, we nonetheless designate these isotopes in parentheses.

The atomic mass of $^{205}$Fr has been previously measured by Penning trap mass spectroscopy \cite{IsolTrap205Fr} to a higher precision than our measurement.  We begin, therefore, by verifying that our result for the atomic mass of $^{205}$Fr agrees with the literature value, as shown in Table~\ref{tbl205FrMass}.  Because our reference, $^{133}$Cs$^+$, is a highly distant reference it is foreseeable that a mass-dependent deviation may be introduced into the evaluation of the atomic mass of $^{205}$Fr as the $^{133}$Cs will experience slightly different radio-frequency fields during its transit from the ion trap. 

\begin{table}[h]
\caption{\label{tbl205FrMass}Mass determination of $^{205}$Fr to validate the measurement method.  The term $\rho$, defined in Eq.~\ref{eqRatio}, is based on $^{133}$Cs$^+$ as a reference.}
\begin{tabular}{cccc}
Laps & $\rho$ & Mass Excess [keV] & Detected Ions \\
\hline
148 & 1.54243949(37) & -1565(47)(60) & 254 \\
147 & 1.54244253(43) & -1188(54)(60) & 210 \\
\hline
Average & 1.54244087(28) & -1390(35)(60) &  \\
\hline
\end{tabular}
\end{table}

Taking the weighted average of the two measurements, we find a mass excess of -1390(35)(60)~keV for $^{205}$Fr, a deviation of $\Delta m$=-80(36)(60)~keV from the values adopted in the most recent atomic mass evaluation (AME2012) \cite{AME2012}; the second uncertainty of 60~keV is a systematic uncertainty from the $t_0$ term in Eq.~\ref{eqRatio}.  While this result is consistent with literature data, we must acknowledge that it cannot be excluded that it may represent a mass-dependent error of $\approx6\times10^{-9}/\Delta u$.  To be as conservative as possible, we therefore use $^{205}$Fr$^+$ as the reference in determining atomic masses of the other RI.  In the case of RI produced with the $^{165}$Ho target, we use the $^{133}$Cs$^+$ as a carrier signal to allow $^{205}$Fr$^+$ to be used as the reference, as described in Eq.~\ref{eqCarrier}:

\begin{align}
\label{eqCarrier}
\rho^\prime=\rho/\rho(^{205}\mathrm{Fr}^+)
\end{align}
where $\rho$ is with respect to $^{133}$Cs$^+$ and $\rho(^{205}\mathrm{Fr}^+)$ is the weighted average value of $\rho$ for $^{205}$Fr$^+$ from Table~\ref{tbl205FrMass}. The results of evaluating the data using $^{205}$Fr$^+$ as the reference are shown in Table~\ref{tabMasses}. 

\begin{table}
\caption{\label{tabMasses}Measured data based on $^{205}$Fr$^+$ references.  The deviation from literature value was calculated as $\Delta m = m_\mathrm{MRTOF} - m_\textrm{lit}$.  See text for details of parenthetical isotopes.}  
\begin{ruledtabular}
\begin{tabular}{c}
$n$=148 Laps
\end{tabular}
\begin{tabular}{cllc}
Species & \ \ \ \ \ \ \ $\rho^\prime$ & Mass Excess [keV] & Detected ions \\
\hline
($^{201}$Bi) & 0.9803821(20)	& -21450(385)(3)	& \,\,\,\,6\\
$^{201}$Po & 0.98040732(39)	& -16642(75)(3)	& \,\,96\\
$^{201}$At & 0.98043596(23)	& -11174(44)(3)	& 531\\
($^{205}$Po) & 0.99990(10)	& -18567(2050)	& \,\,\,\,2\\
($^{205}$At) & 0.999939(22)	& -12898(420)		& \,\,11\\
($^{205}$Rn) & 0.999967(16)	& \,\,-7502(320)		& \,\,40\\
($^{206}$At) & 1.004819(11)	& -12497(2150)(1)	& \,\,\,\,3\\
($^{206}$Rn) & 1.0048400(30)	& \,\,-8565(600)(1)	& \,\,\,\,3\\
$^{206}$Fr & 1.00487949(61)	& \,\,-1043(120)(1)	& 133\\
\end{tabular}
\begin{tabular}{c}
$n$=147 Laps
\end{tabular}
\begin{tabular}{cllc}
Species & \ \ \ \ \ \ \ $\rho^\prime$ & Mass Excess [keV] & Detected ions \\
\hline
$^{201}$Po & 0.98041151(27)		& -15842(521)(3)	& 11\\
$^{201}$At &  0.98044042(43)		& -10321(820)(3)	& \,\,4\\
$^{206}$Fr & 1.0048777(13)		& \,\,-1371(250)(1)	& 66\\
\end{tabular}
\begin{tabular}{c}
Weighted average
\end{tabular}
\begin{tabular}{clll}
Species & \ \ \ \ \ \ \ $\rho^\prime$ & Mass Excess [keV] & \,\, $\Delta m$ [keV] \\
\hline
($^{201}$Bi) &0.9803821(20)				& -21450(385)(3)	&  \, \, -35(385)(3)\\
$^{201g}$Po & 0.98040785(39)			& -16541(74)(3)	&  \, \, -16(74)(3) \\
$^{201}$At & 0.98043619(23)				& -11131(44)(3) 	& \, -342(45)(3) \\
($^{205}$Po)  & 0.99990(10)				& -18567(2050)	& -1059(2050)\\
($^{205}$At) & 0.999939(22)				& -12898(420)		& \ \, \, 73(420)\\
($^{205}$Rn) & 0.999967(16)				& \,\,-7502(320)	& \ \, 212(320)\\
($^{206}$At) & 1.004819(11)				& -12497(2150)(1)	& \ \ \ -69(2150)(1)\\
($^{206}$Rn) & 1.0048400(30)				& \,\,-8565(600)(1)	& \ \ \ 551(600)(1)\\
$^{206m}$Fr$^\dagger$ & 1.00487892(55)	& \,\,-1150(107)(1)	& \ \ \ -98(118)(1)\\
\end{tabular}
$^\dagger$ \tiny{May be $^{206g}$Fr, with deviation from literature of $\Delta m$=92(111)(1)~keV}
\end{ruledtabular}
\end{table}

The weighted average data is generally in agreement with AME2012 literature values.  Penning trap data for $^{205,206}$Rn \cite{SHIPTrapRn} have been reported since the most recent AME2012; our values are consistent with these values.  

In the case of $^{201}$Po, which is known to have an isomeric state ($T_{1/2}$=8.96 min) with excitation energy $E_\mathrm{Ex}$=424~keV, we presume the ions originate from $\beta$-decay of $^{201}$At which had settled upon a surface within the gas cell other than the RF-carpet and were ejected from the surface by decay recoil.  From spin considerations, we can assume the $\beta$-decay would primarily populate the ground state. The AME2012 mass uncertainties for $^{201}$Po and $^{201}$At are dominated by ESR data \cite{ESR_Po_At}.  Assuming $^{201g}$Po, we find reasonable agreement with literature, but find a large deviation in the case of $^{201}$At.

Previous atomic mass values for $^{205, 206}$At and $^{206}$Fr were limited to those derived from $\alpha$-decay studies.  In the cases of $^{205, 206}$At we find good agreement with the literature.  In the case of $^{206}$Fr, two long-lived isomeric states with sufficient half-lives for our experiment to observe have been reported \cite{92Hu04}. The lower-lying of these two isomers is listed as having an excitation energy of $E_\mathrm{Ex}$=190(40)~keV and spin (7+) while the higher-lying isomer has $E_\mathrm{Ex}$=730(40)~keV and spin (10-). From our data we cannot differentiate between the ground state and this first isomeric state, but find consistency with a mixture.  It is expected that the isomeric yield should scale as $Y_m/Y_g\sim (2J_m+1)/(2J_g+1)$, where $J_m$ and $J_g$ are the spin of the isomeric and ground state, respectively \cite{IsomericRatio}.  Using peak amplitude as a proxy for yield, we performed two-peak fitting with the inter-peak separation fixed at 15~ns (corresponding to 730~keV) and extract a yield ratio of $Y_{n}/Y_{g,m}$=0.09(14); the spin assignments would predict $Y_n/Y_{g,m}\approx$1.0.  This indicates that either the reaction imparted insufficient angular momentum to populate the (10-) isomer or that that isomer has a spin lower than that of the (7+) isomer.  In the former case, the (7+) isomer is likely to be similarly underproduced and our data should be dominated by the ground state nucleus.

These results demonstrate the potential power of the MRTOF-MS technique.  We have demonstrated the semi-unique ability to perform simultaneous mass measurements of isobar chains.  In the future, the ability to simultaneously measure long isobar chains will allow the MRTOF-MS to make a powerful impact on the mass landscape by applying the technique to in-flight fission and fragmentation beams.  Additionally, we demonstrate an ability to directly measure isomeric production ratios by mass spectroscopy.  Perhaps most importantly, we demonstrate the ability to utilize low statistics to identify ions.  The ability to achieve ppm-level measurements with $\lesssim$10~detected ions will soon allow the device to be applied to trans-Uranium RI beams, eventually leading to identification of new SHE via mass spectroscopy.

We wish to express gratitude to the Nishina Center for Accelerator Research and the Center for Nuclear Science at Tokyo University for their support of online measurements.  This work was supported by the Japan Society for the Promotion of Science KAKENHI (Grant Numbers 2200823, 24224008, 24740142, and 15K05116).

\end{document}